# Dynamic Scheduling and Workforce Assignment in Open Source Software Development

**Hui Xi, Dong Xu and Young-Jun Son**
**Systems and Industrial Engineering, The University of Arizona**
**Tucson, AZ, USA**

A novel modeling framework is proposed for dynamic scheduling of projects and workforce assignment in open source software development (OSSD). The goal is to help project managers in OSSD distribute workforce to multiple projects to achieve high efficiency in software development (e.g. high workforce utilization and short development time) while ensuring the quality of deliverables (e.g. code modularity and software security). The proposed framework consists of two models: 1) a system dynamic model coupled with a meta-heuristic to obtain an optimal schedule of software development projects considering their attributes (e.g. priority, effort, duration) and 2) an agent based model to represent the development community as a social network, where development managers form an optimal team for each project and balance the workload among multiple scheduled projects based on the optimal schedule obtained from the system dynamic model. To illustrate the proposed framework, a software enhancement request process in Kuali foundation is used as a case study. Survey data collected from the Kuali development managers, project managers and actual historical enhancement requests have been used to construct the proposed models. Extensive experiments are conducted to demonstrate the impact of varying parameters on the considered efficiency and quality.

## 1. Introduction
In a traditional software development process, project scheduling and workforce assignment take place sequentially. This approach, however, is not suitable for an open source software development (OSSD) process, where the system condition changes frequently over time such as workforce, project extension, product rework and other unexpected events. In this paper, a novel simulation-based framework is proposed for integrating dynamic project scheduling and workforce assignment in OSSD. The proposed framework intends to reduce the project waiting time as well as to maximize workforce utilization by conducting the project scheduling and workforce assignment simultaneously.

Many efforts have been made in the literature on exploring the project scheduling and workforce assignment problems. Heimerl and Kolisch [1] proposed a mixed-integer linear program (MIP) with a tight LP-bound to solve project scheduling and multi-skilled resources assignment problems simultaneously. Different problem parameter setups such as the project window size, number of resources, and workload are constructed for comparison purpose, and the results have shown that the proposed MIP can reduce cost tremendously compared with other simple heuristics used in practice. Valls et al. [2] proposed a hybrid genetic algorithm to solve the task scheduling and resource assignment problems with multiple constraints considered such as penalties for project delivery delay, client priorities, and precedence relationship. The results revealed that the proposed hybrid genetic algorithm is effective and efficient. Later, Tiwari et al. [3] used an integer programming optimization procedure to solve a multi-mode, resource-constrained, project scheduling problem (MRCPSP). The results revealed that the output of the model can help to increase throughput (reducing make-span) by identifying resource bottlenecks, and cross-training proper groups or individuals. Lee et al. [4] has proposed a system dynamics model, which can deal with the project scheduling problem considering project properties, resource availability as well as system functionality; Celik et al. [5] proposed a simulation-based workforce assignment model considering developer embeddedness in a social network.

More recently, research has been conducted in various aspects for successful OSS development. Toral et al. [6] focused on the social relationships among community members, where they analyzed different members' roles and their impact on success in the OSS development. Results in their research revealed the importance of selecting active developers and core members in the development of virtual communities. However, since this work did not utilize the social network variables sufficiently, Toral et al. [7] did another research on the collaborative behavior of virtual communities involving extensive social network analyses. Results from this paper have demonstrated the importance of brokers in helping OSS projects to engage in a discourse and co-learning experience with their user communities, and thus improving the efficiency of the OSS development process. Besides the efficiency of the development process, other criteria of OSS development have also drawn lots of researchers' attention. Later, Capra et al. [8] investigated the relationship among software design quality, development effort, and governance in OSSD. They found out that contributors tend to increase their efforts if they are under less pressure from deadlines. Also, they provided empirical evidences supporting their hypothesis that less formal governance at the end of the development life cycle facilitates coordination among developers and thus to insure a higher level of code quality.

The remainder of this paper is organized as follows. In Section 2, a detailed discussion on the two level simulation-based dynamic scheduling and workforce assignment framework proposed in this paper is given. Also, a case study involving enhancement request process of Kuali Foundation is discussed. In Section 3, the proposed framework is illustrated and demonstrated with the considered case study. Finally, conclusions and future work are presented in Section 4.

## 2. Simulation-based Dynamic Scheduling and Workforce Assignment Framework

In this work, we propose a two-level modeling framework to deal with project scheduling and work for assignment in OSSD, respectively. A system dynamic (SD) simulation model is constructed to devise a balanced schedule for multiple ongoing projects based on their specific requirements (e.g. deadline, minimum workload and skill required, deliverable type). The devised project schedule will then be used as an input to an agent-based (AB) simulation model to devise an optimal workforce assignment among these projects considering both developers' expertise and workload. To illustrate the proposed framework, an enhancement process in Kuali Foundation has been adopted as a case study. Section 2.1 describes a brief background of Kuali Foundation, its organization structure, and the enhancement request process. In Sections 2.2 and 2.3, an SD-based scheduling model and an AB workforce assignment model will be illustrated in a greater detail. Section 2.4 discusses the interaction issues between the proposed SD and AB models.

### 2.1 Kuali Organization Structure and Enhancement Request Process

Kuali Foundation is a nonprofit community for OSSD for higher education. The members of the Kuali include organizations from both academia and industry and have been partnered to deliver distributed, modular and open-source software for all sizes of institutions. Figure 1 depicts an overview of the functional management structure of the Kuali Foundation. The enterprise software systems consist of Kuali Financial System (KFS), Kuali Coeus (KC) for research administration, and Kuali Student (KS) (http://kuali.org/about). In this work, we focus on the Kuali Financial System (KFS). As Kuali's initial product, KFS aims at bringing the proven functionality of legacy applications the ease and universality of online services. KFS meets the needs of all Carnegie Class institutions, and its flexible and modular structure allows the systems to be scaled to different complex level so as to meet the need of various organizations.

In Kuali community, when a participating institution needs software development or maintenance, it has to require the service through the enhancement request process. The process start with submission of an enhancement request form by a Subject Matter Expert (SME) from the institution. The form contains the following required information: 1) the primary component involved (KFS module in this case), 2) the priority (high, medium and low), 3) business need, 4) impact if not implemented, and 5) questions to be answered by the Development Manager (DM). Once receiving the form, DM will then add more information regarding this particular request, including estimated effort required, technical impact (required technical challenge), and system impact (expected functional impact on any other modules or areas within the system). Next, the form will be submitted to the Functional Council (FC) which then makes the final decision on whether to accept this request. If the request is approved by the FC, the FC meets with the Project Manager (PM) in order to schedule the enhancement and delivers this decision to the DM. However, if the request is rejected, the SME may choose to appeal to the board regarding the decision. After reviewing the appeal, the board may affirm or override the FC's decision. As soon as the request is accepted, the FC will start scheduling the new project with other on-going projects.

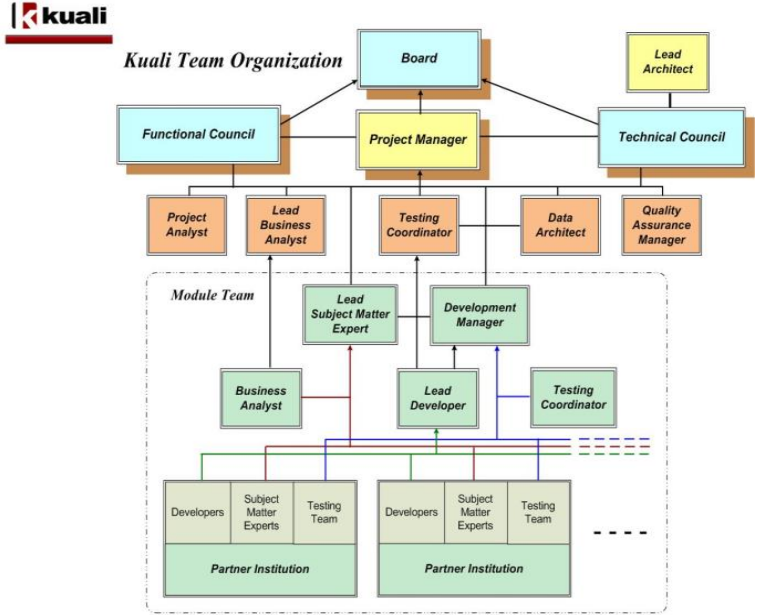

Figure 1: Kuali team organization

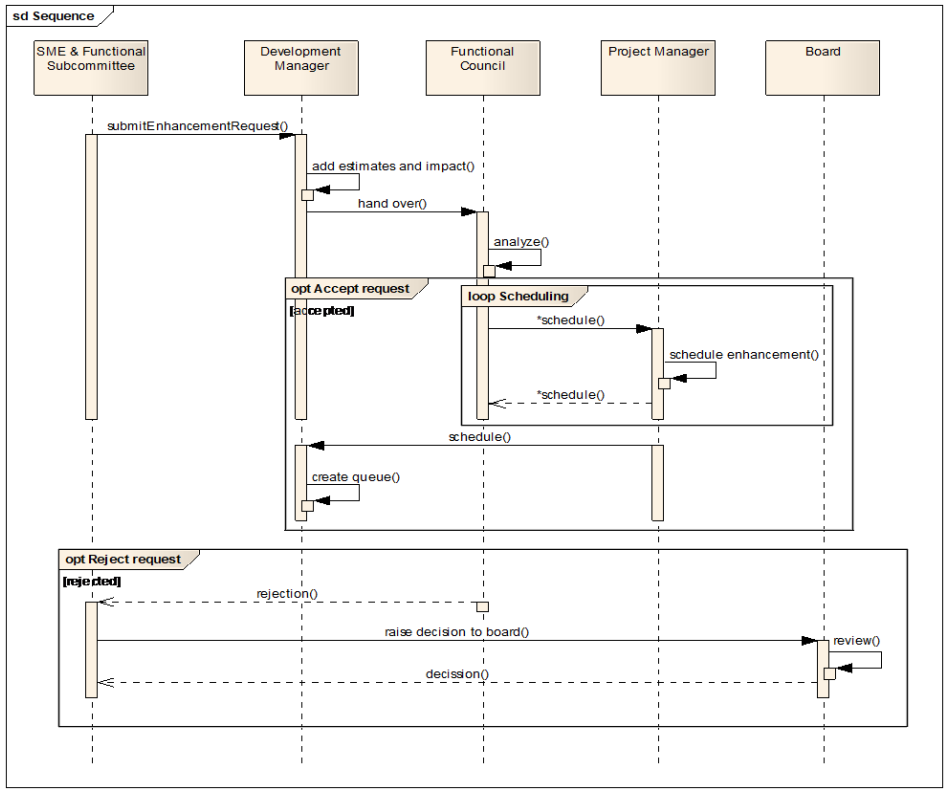

Figure 2: Sequence diagram for enhancement request process

## 2.2 System Dynamics Model for Project Scheduling

In this section, we propose to use a system dynamic (SD) model to represent the project scheduling process. As shown in Figure 2, Functional Council (FC) and Project Manager (PM) work hand in hand to achieve an agreed schedule for all the accepted projects. In the scheduling process, FC has to take into account of multiple attributes of each project and make sure that every project meets the timeline as well as maintains a high level of product quality. When a new project comes in, FC will evaluate it considering four attributes: Project deadline, Institution priority, Estimated Effort, and Required Expertise. At the same time, FC checks with PM to find the currently available workforce level and their expertise. The causal loop diagram in Figure 3 depicts how these attributes would affect the start time of each project, where each node is a variable and each link denotes the relationships between the nodes. A solid line represents a positive relationship (e.g., an increase in node A results in an increase in node B), while a dotted line refers to an inverse relationship (an increase in A decreases B). As shown in Figure 3, a project requested by an institution with a higher priority may start earlier. However, if a project requires workforce with particular expertise, it may start only when those developers are available. Similarly, a project will be scheduled later if it requires more efforts or its deadline is later.

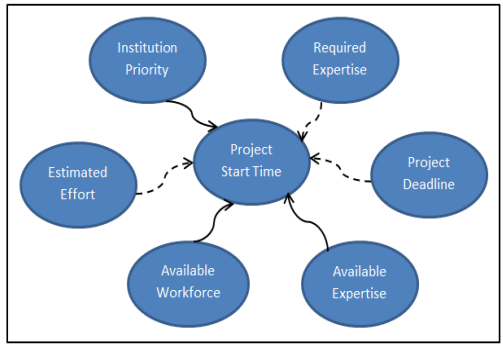
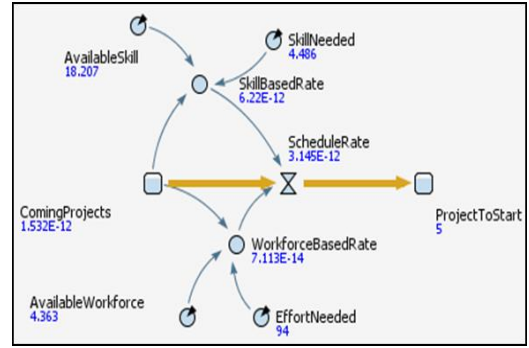

Figure 3: Causal loop diagram for project scheduling process

Figure 4: Stock-flow diagram for project scheduling process

Figure 4 depicts the proposed system dynamic model for the considered project scheduling based on the causal loop diagram in Figure 3. Newly coming projects are stored in a FIFO queue and will be scheduled with a rate affected by the currently available workforce and available expertise (see Equations (1-3)). We demote the total number of coming projects as $N(t)$. The system keeps checking the available workforce and available skill on daily basis and updates the number of project that can be scheduled. Once we obtain the number of projects that can be scheduled, we retrieve those many of projects from the FIFO queue. Then, we reorder them by the proposed algorithm shown in Figure 5.

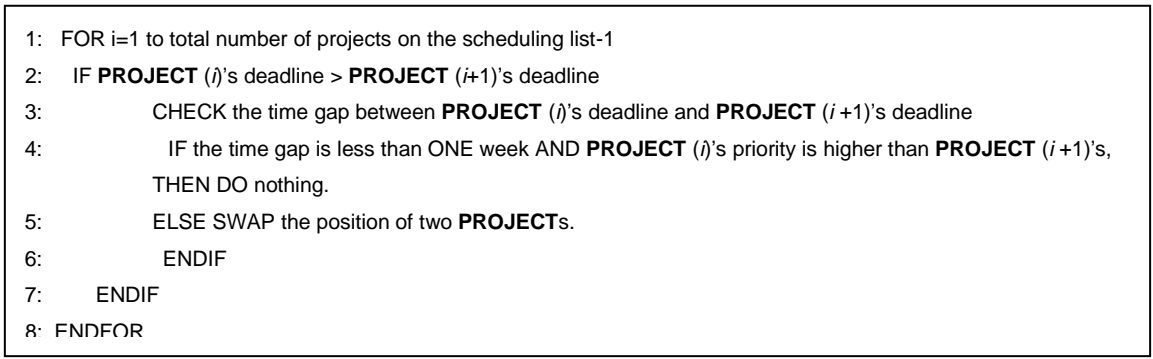

Figure 5: Pseudo code of the proposed scheduling algorithm

$$ScheduleRate(t) = \frac{WorkforceBasedRate(t) + SkillBasedRate(t)}{2} \qquad (1)$$

$$WorkforceBasedRate(t) = \frac{AvailableWorkforce(t)}{\sum_{i=0}^{N(t)} EstimatedEffort_i} N(t) \quad (2) \qquad SkillBasedRate(t) = \frac{AvailableSkill(t)}{\sum_{i=0}^{N(t)} ExpertiseLevel_i} N(t)$$

(3)

### 2.3 Agent-based Simulation for Workforce Assignment

In this section, we will discuss the workforce assignment issues using the agent based simulation technology. Multiple objectives need to be considered when conducting the optimal workforce assignment. First, efficiency of the OSSD process is concerned for the benefits of organizations, which involve issues such as workforce load balancing and software development time. Second, the quality of deliverables is also concerned for customer benefits, where the quality metrics include software property, code modularity, and the number of errors. In this work, we propose to conduct a workforce assignment via multi-objective optimization considering the above mentioned two sets of objectives (organizational benefits and customer benefits). The survey data that we have collected indicated that the quality of deliverables (e.g. code modularity, less numbers of bug and potential error) depend highly on the multi-skilled, work load balanced team formation. Figure 6 provides the overview of workforce assignment process.

First, the Putnam SLIM Model [9] is used to estimate the number of employees needed for each project, where the SLIM Model is based on Putnam's analysis of the software life cycle in terms of the Rayleigh distribution of project personnel level versus time. The macro-estimation model for the effort used in SLIM is

$$S_s = C_k K^{(1/3)} t_d^{(4/3)} \quad (4)$$

where $S_s$ is the number of delivered source code and instructions; $K$ denotes lift-cycle effort in person-years; $t_d$ is the development time in years; $C_k$ is a "technology constant". In our work, the estimated number of delivered source code and instructions has been estimated based on the similar software enhancement request received before. The development time has been set from the workforce assignment starting time to the due date of the enhancement request. Values of the "technology constant" typically range between 610 and 57,314. This version of SLIM model can estimate $C_k$ as a function of various variables such as project's use of modern programming practices, hardware constraints, personnel experience as well as interactive development activities.

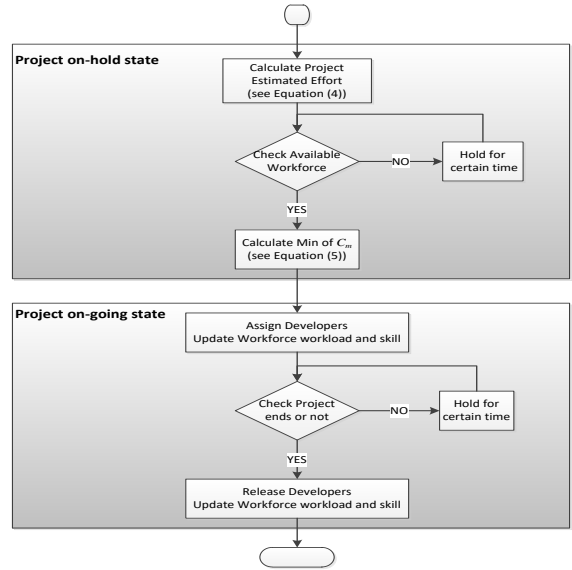

Figure 6: Flowchart of workforce assignment process

Second, based on the required employee participation estimated from the first step, we identify the corresponding mixes of employees considering both the currently available workforce together with their workload as well as their skill set. It is noted that the workforce workload is dynamically changing during the simulation run. In terms of workforce skill set, different employees possess different levels (e.g. good and moderate) skills, and they impact the quality of the final software as well as the completion time. According to the survey data that we have obtained, projects that ended up with a high quality have some commonality on the team formation. It was also observed that the projects with different goals (e.g. time urgency versus quality oriented) have different emphases on the required employee skills for the optimal team formation. The workforce skill values for each employee are captured by the survey data; different ranges of skill value represent different degrees of matches to a certain type of project. In this work, we will devise the workforce combination based on the equation (5) shown below.

$$C_m = \min_n \left\{ \sum_i^K (\alpha_m f(w_i(n)) + \beta_m g(s_i(n))) \right\} \quad \text{where} \quad n = \binom{46}{K} \quad (5)$$

$\alpha$ and $\beta$ provides the weights (relative importance) of workforce load and workforce skill in project *m*, respectively; $w_i(n)$ denotes the workload of developer *i* at the $n^{th}$ team combination, $s_i(n)$ denotes the skill of developer *i* at the $n^{th}$ team combination; *f* and *g* function provides the mapping between the developer *i*'s workload ($w_i$) or skill ($s_i$) to the influence of overall team formation; *m* is the project number; *n* represents all the possible workforce team combinations, *K* is the estimated effort (number of developers required); $C_m$ gives the optimal workforce team combination. All *n* combinations of project team formation are calculated, and then the lowest value of $C_m$ is selected.

Third, after the optimal $C_m$ is selected, the scheduled project(s) go from the on-hold state to the on-going state (see Figure 6). At the project on-going state, workforce assignment will be conducted and workforce workload and skill will be updated. During the simulation run, the system checks the project status (e.g. finished or not) on daily basis until the project ends. Then, corresponding workforce will be released and the workforce workload and skill set will be updated.

## 2.4 Interaction between Dynamic Scheduling (System Dynamics) and Workforce Assignment (Agent based Simulation)

Figure 6 depicts the interactions between the proposed dynamic schedule ng model (SD model) and the workforce assignment model (AB model). Two types of events exist during the scheduling process, where the first one generates projects, and the second one schedules those projects. In this work, it is assumed that projects are generated based on an exponential distribution, where the mean value has been obtained from the historical enhancement request data available from Kuali. The event of scheduling project is triggered by one of the following two events: 1) a new project is generated; or 2) workforce availability is updated. The scheduling project event, which implements the algorithm in Figure 5, updates the project order in the queue based on the current system condition (e.g. the number of projects waiting and ongoing, current workforce availability considering both workload and skill). Whenever the *ProjectToStart* variable (stock) in Figure 4 reaches an integer number *n* (*n*=1,2,…), the corresponding *n* projects at the waiting queue will start to assign developers. Then, agent-based workforce assignment model will be triggered by the newly started project(s). The workforce workload and skill set will be updated in the following two cases: 1) the process of assigning workforce occurring at the project starting time; 2) the process of releasing workforce occurring at the project ending time. The schedule rate e.g. *SkillBasedRate*, *WorkforceBasedRate* in the system dynamics model (see Figure 4), which controls the project scheduling process, is adaptively updated by the updating process at the agent-based model. The above mentioned interactions are continued until all projects are finished.

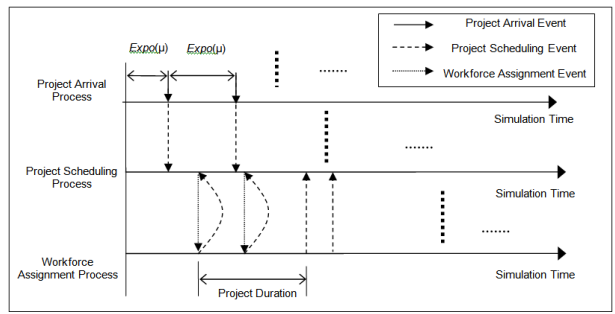

Figure 6: Interaction overview of dynamic scheduling process and workforce assignment process

## 3. System Implementation and Experiments

## 3.1 System Implementation

The proposed framework has been implemented on commercial multi-paradigm simulation software, AnyLogic. Figure 7 shows major functions, parameters and databases used to build the proposed system dynamics (SD) model as well as the agent-based simulation (ABS) model. Two databases are used; one is project database (*ProjectDB*, see Figure 7), which stores all the information about projects such as project ID, project deadline, priority, estimated effort, and required expertise; the other one is workforce database (*SkillDB*, see Figure 7), which saves all the information about workforce such as employee technical skill, Kuail experience, and leadership level. Two types of events (*Generate_Project*, *Check_Scheduling*) have been implemented in the project scheduling model, and the workload and skill variables (e.g. *SkillBasedRate*, *WorkforceBasedRate*) are the interface used for communications between the SD and ABS models, which have been discussed in Section 2.4.

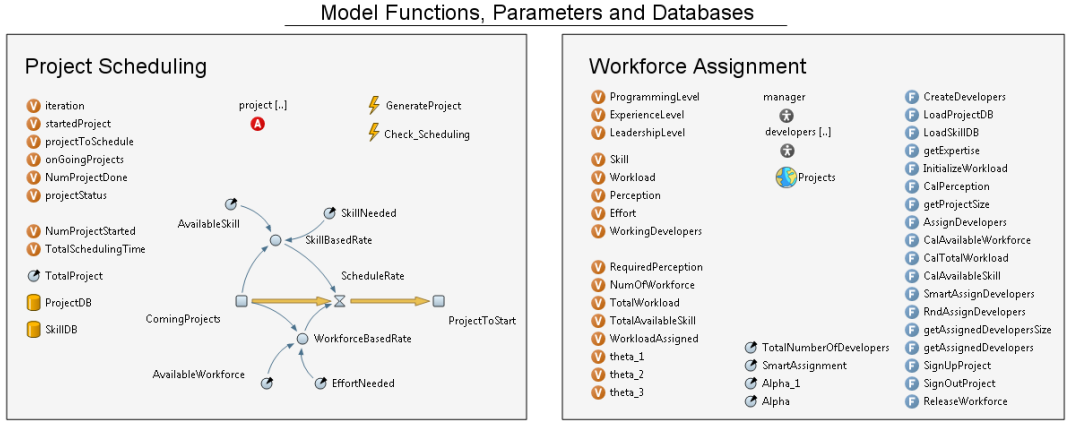

Figure 7: Overview of system implementation

## 3.2 Experiments and Results

In this section, the constructed models conforming to the prosed framework are illustrated and demonstrated for the enhancement request process in Kuali. The total number of projects and developers considered are 30 and 46, respectively. The range of project duration is set from 0.3 to 0.7 years. The overall simulation length is set as 3 years, and the time unit is day.

Three metrics have been chosen to represent system performance, which include average project waiting time, average workforce utilization, and average project queue length. Table 1 provides comparison results for the same system setups obtained from two different approaches: the proposed SD and ABS integrated framework and a method without using the proposed framework (using FIFO rule). As shown in Table 1, the average project waiting time is reduced from 80.1486 to 46.3644 (for a total 30 projects), which gives 38.41% improvement. The other two criteria have been also improved, involving 14.59 % improvement on the average workforce utilization and 28.02% improvement on the average project queue length.

Table 1: Results comparison using proposed framework with initial solution (10 replications & 95% C.I.)

|  | Average project waiting time | Average workforce utilization | Average project queue length |
|---|---|---|---|
| Without using proposed framework | 80.1486+12.3681 | 0.3777+0.0292 | 2.6763+0.4082 |
| Using proposed framework | 49.3644+11.8553 | 0.3226+0.0244 | 1.9273+0.2973 |

Figure 8(a) compares the project average waiting time obtained from the proposed framework and the system with a FIFO scheduling rule over 10 simulation replications. As shown in figure 8(a), the proposed framework almost always outperforms the reference case except one replication. Figure 8(b) provides similar comparisons for the workforce average utilizations. While the improvement in this case is not as significant as the case with the project average waiting time, it reveals that even with the same workforce assignment approach, different project scheduling methods can result in different workforce utilization. Under the same problem setups (e.g. the number of developers, the number of projects), an approach involving a lower workforce utilization will be a more efficient one.

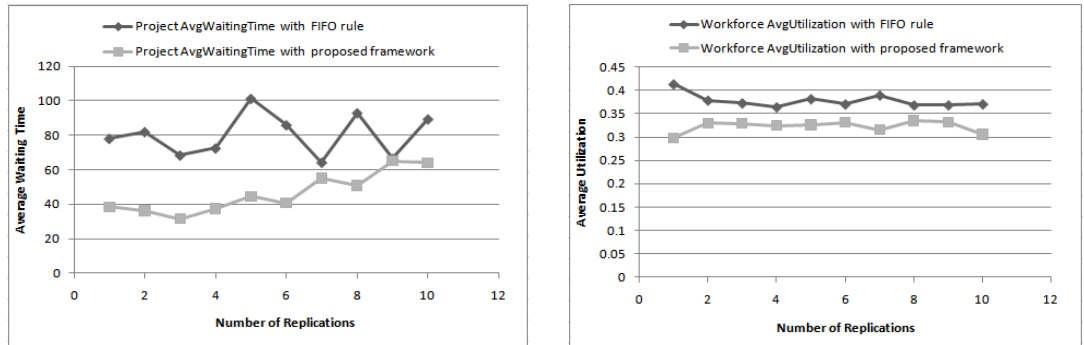

Figure 8: (a) Project average waiting time comparison for 10 simulation replications; (b) Workforce average utilization comparison for 10 simulation replications

## 4. Conclusion and Future Work

In this paper, a hybrid simulation-based framework has been proposed for integration of dynamic project scheduling and workforce assignment. Under the proposed framework, a system dynamics (SD) model has been developed for project scheduling, and an agent-based (AB) model has been developed for workforce assignment. Interaction between those two simulation models have been discussed in detail. To illustrate and demonstrate the utility of the proposed framework, enhancement request processes at Kuali organization have been used as a case study. The results obtained from various experiments revealed that the proposed integrated framework is more effective as well as efficient compared with other traditional methods. As part of the future work, different problem setups (e.g. variations in total number of projects, number of developers, arrival rate, and project duration) will be considered and analyzed. In addition, more detailed scheduling heuristics available in the literature will be tested compared with the dynamic scheduling rule in the proposed framework. Finally, other approaches will be devised for workforce assignment (where the dynamic scheduling model is fixed), which will be compared with the current approach considering workforce workload and skill set.

## Acknowledgements
This work was supported by the National Science of Foundation under NSF-SOD 0725336.

## References
1. Heimerl, C., and Kolisch, R., 2010, "Scheduling and staffing multiple projects with a multi-skilled workforce," OR Spectrum, 32(2), 343-368.
2. Valls, V., Perez, A., Quintanilla, S., 2009, "Skilled workforce scheduling in Service Centres," European Journal of Operational Research, 193(3), 791-804.
3. Tiwari, V., Patterson, J.H., and Mabert, V.A., 2009, "Scheduling projects with heterogeneous resources to meet time and quality objectives," European Journal of Operational Research, 193(3), 780-790.
4. Lee, S., Celik, N., Son, Y., 2009, "An Integrated Simulation Modeling Framework for Decision Aids in Enterprise Software Development Process," International Journal of Simulation and Process Modeling, 5(1), 62-76.
5. Celik, N., Lee, S., Mazhari, E., Son, Y., Lemaire, R., and Provan, K., 2009, Simulation-based Workforce Assignment in a Multi-organizational Social Network for Community-based Software Development (submitted).
6. Toral, S.L., Martínez-Torres, M.R., and Barrero, F., 2009, "Virtual Communities as A Resource for the Development of OSS Projects: The Case of Linux Ports to Embedded Processors," Behaviour & Information Technology, 28(5), 405-419.
7. Toral, S.L., Martínez-Torres, M.R., and Barrero, F., 2010, "Analysis of Virtual Communities Supporting OSS Projects Using Social Network Analysis," Information and Software Technology, 52(3), 296-303.
8. Capra, E., Francalanci, C., and Merlo, F., 2008, "An Empirical Study on the Relationship among Software Design Quality, Development Effort, and Governance in Open Source Projects," IEEE Transactions on Software Engineering, 34(6), 765-782.
9. Putnam, L.H., 1978, "A General Empirical Solution to the Macro Software Sizing and Estimating Problem," IEEE Transactions on Software Engineering, SE-4(4), 345-361.